\documentclass[letterpaper]{article}
\usepackage{natbib,alifeconf}
\usepackage{algorithm2e}
\usepackage{graphicx}
\usepackage{amsmath}
\usepackage{amssymb}
\usepackage{caption}
\usepackage[font=small]{caption}
\usepackage{mwe}
\usepackage{hyperref}
\graphicspath{ {images/} }
\usepackage{subfigure}
\usepackage{xcolor}
\usepackage[normalem]{ulem}

\title{CIMAX: Collective Information Maximization in Robotic Swarms Using Local Communication}
\author{Hannes Hornischer$^{1,2,}$\thanks{\;\;Corresponding author: hannes.hornischer@edu.uni-graz.at}\;\,, Joshua Cherian Varughese$^{1,3}$, \\ {\Large Ronald Thenius$^{1}$, Franz Wotawa$^{3}$, Manfred F{\"u}llsack$^{2}$ and Thomas Schmickl$^{1}$}
\mbox{}\\
$^1$Artificial Life Laboratory, Department of Zoology, Institute of Biology,
Karl-Franzens-University, Graz, Austria \\
$^2$Institute of Systems Sciences, Innovation and Sustainability Research, Karl-Franzens-University, Graz, Austria \\
$^3$Institute for Software Technology, Technical University of Graz, Austria }

\begin{document}
\maketitle


\begin{abstract}

Robotic swarms and mobile sensor networks are used for environmental monitoring in various domains and areas of operation. Especially in otherwise inaccessible environments decentralized robotic swarms can be advantageous due to their high spatial resolution of measurements and resilience to failure of individuals in the swarm. However, such robotic swarms might need to be able to compensate misplacement during deployment or adapt to dynamical changes in the environment. Reaching a collective decision in a swarm with limited communication abilities without a central entity serving as decision-maker can be a challenging task. Here we present the CIMAX algorithm for collective decision making for maximizing the information gathered by the swarm as a whole. Agents negotiate based on their individual sensor readings and ultimately make a decision for collectively moving in a particular direction so that the swarm as a whole increases the amount of relevant measurements and thus accessible information. We use both simulation and real robotic experiments for presenting, testing and validating our algorithm. CIMAX is designed to be used in underwater swarm robots for troubleshooting an oxygen depletion phenomenon known as ``anoxia''.

\end{abstract}

\section{Introduction\label{sec:intro}}

Swarms of various lifeforms have been observed to utilize emergent group dynamics~\citep{eberhart2001swarm} for various tasks such as foraging~\citep{Seeley92}, reproduction~\citep{bonner1949social, durston1973dictyostelium} or escaping predators~\citep{cavagna2010scale, brock1960fish, magurran1987provenance}.
~\cite{Seeley92} discovered how bees use waggle dances for foraging by pointing their hive to high quality food sources. 
~\cite{bonner1949social} and~\cite{durston1973dictyostelium} examined the communication behavior of slime mould cells which despite its simplicity enables self-organization with respect to foraging, reproduction et cetera.
~\cite{cavagna2010scale} analyzed how starling flocks respond to external stimuli as a collective in order to evade predators.
Due to the availability of many eyes in a swarm, each individual spends less time on scanning the area for predators while spending more time on foraging. \cite{magurran1987provenance} experimentally demonstrated various formations used by shoals of minnows when detecting predators. Decentralized intelligence of such kind is popularly known as swarm intelligence~\citep{beni89}.
Natural systems exhibiting this decentralized intelligence have inspired researchers due to their adaptability to the environment, resilience to perturbations and underlying simplicity. However despite simple rules governing the behavior of individuals in a swarm, the resulting collective behavior often shows a stunning degree of complexity -- as it can also be observed in the synchronized flashing of the fireflies {\it lampyridae}~\citep{buck1988synchronous}.
Researchers in many emerging fields such as ubiquitous computing~\citep{ubiswarm}, multi-robot systems~\citep{zahadat2016division, kernbach2009re}, traffic management~\citep{renfrew2009traffic} etc. have recognized parallels between such multi-agent artificial systems and natural systems containing several actors~\citep{Garnier2007}. 
As a result, extensive research has been dedicated to self organization and decentralization in complex systems~\citep{Dorigo2004}. When designing swarms or sensor networks one challenge that often needs to be addressed is the collective decision making task~\citep{Kernbach01012013}.
In this paper we present an algorithm enabling a swarm of individuals with limited communication abilities to make a collective decision regarding its direction of motion in order to maximize information accessible to the collective.

The algorithm presented in this paper enables a swarm to increase its information entropy over time. For example consider a swarm of $N$ agents and each measurement $x_{i}$ independently follows a uniform probability distribution, ${p(x_{i}) \ = \ 1/M }$ with $M$ possible measurements. The resulting information entropy for each agent according to Shannon's measure of information entropy 
\begin{equation}
\label{eq:shannon}
H(x) \ = \  -\sum_{i=1}^{M}p(x_{i})log_{b}(p(x_{i}))
\end{equation}
for a binary system is ${H  =  log_2(M)}$. The entropy for $N$ independent agents is ${H = N \cdot (log_2(M))}$. As the number of possible options in the distribution decreases, the information entropy in the entire system decreases. In terms of the quantities measured by the swarm, for larger variance in those measurements we have larger values of $M$ and therefore larger information entropy of the swarm. \\
In the algorithm presented in this work we use the variance in measurements of the swarm in combination with a simple bio-inspired communication mechanism to enable swarms to maximize the information available to them. Swarms move in a direction which leads to an increase in information available to the swarm as a whole. In contrast to centralized swarms here the individual agents use only local information.
In the following we refer to the algorithm as CIMAX.

We initially designed CIMAX to address the task of documenting, examining and ultimately forecasting the frequently but irregularly appearing anoxic waters phenomena~\citep{Anoxia} in the lagoon of Venice. 
During this phenomenon which we refer to as ``anoxia'', the oxygen content of a small part of the lagoon decreases dramatically resulting in the death of animal life in that specific area. Anoxia adversely affects the flora and fauna in the lagoon and also causes difficulties for the inhabitants and tourists in Venice.\\
A strategy to examine and document this phenomenon is to utilize a swarm of underwater robots for monitoring a set of environmental parameters, i.a. oxygen concentration levels. For determining dynamics and spatio-temporal evolution of anoxic areas the water body a swarm of robots allows monitoring at various underwater locations and thus high spatial resolution.
One implementation of such swarm robots used for autonomous long-term underwater monitoring was developed and extensively tested in real-world marine environments within project ``subCULTron''~\citep{subCULTron}: the so-called ``aMussel''~\citep{donati2017amussels}.\\
Due to problem such as expensive hardware~\citep{akyildiz2005underwater}, high power consumption~\citep{stojanovic2009underwater} and general complexity of communication underwater~\citep{lanbo2008prospects} the main approach for communication within members of a swarm of aMussels is based on using modulated light for local information transfer.\\

When deploying a swarm at a target location it is not guaranteed that the location is sufficiently covered. It is possible that only few robots are in contact with the anoxic area and the majority of the swarm is not. 
Moreover, even in case the swarm is optimally placed the target area is dynamic and hence mobile. 
Therefore, the CIMAX algorithm can continuously guide the swarm to areas of interest.

The problem that CIMAX addresses in this paper is a classical problem of collective decision making in multi agent systems where individual entities might make conflicting decisions based on local information. According to~\cite{Trianni2015}, algorithms for collective decision making in natural and artificial swarms can be categorized into three main mechanisms.
In the first mechanism, the swarm waits for one entity to have enough information to make a decision and then propagate that decision within the swarm. Organizational structures following this mechanism can be found in form of hierarchies within animal and human societies~\citep{rabb1967social,ahl1996hierarchy}.
The second mechanism is called opinion averaging in which all individuals constantly adjust their own opinion based on their neighbours' opinions until the entire swarm eventually converges to one opinion. This mechanism for collective decision making in robots swarm can also be found in groups of animals which use it for effectively navigating as a collective~\citep{simons2004many,codling2007group}. 
The third mechanism is based on amplification of a particular opinion to produce a collective decision. In this mechanism, each individual randomly starts with an opinion and then changes their opinion to other opinions depending on how often they hear the latter opinion. The amplification mechanism is also found within animals such as the pheromone trails selection in ants~\citep{beckers1990collective} or the temperature based site selection of young bees~\citep{szopek2013}. 
The underlying mechanism of collective decision making of the algorithm presented in this paper relies on the amplification of the mostly held opinions within the swarm which is associated with the second category of mechanisms presented in~\citep{Trianni2015}.

Apart from collective decision making in swarm robotics, our approach is broadly related to relocation of sensor nodes in mobile wireless sensor networks (``MWSN'')~\citep{wang2005sensor, li2007mesh, cui2004swarm}. When deploying a swarm e.g. in an otherwise inaccessible environment, the swarm is often not arranged properly for effective measurement and observation due to inaccurate knowledge of target area, of dynamic changes in local conditions or of unforeseeable events. For optimizing parameters such as coverage, connectivity or network longevity individual members of the network need to be relocated for which a variety of approaches has been suggested.
 
While some approaches in sensor relocation rely on having access to global information~\citep{wang2005sensor} of the position of sensors, often this problem is approached in a decentralized manner. In~\citep{wang2005sensor} the exact position of the sensors is known by the base station or similar central entity. The area covered by the sensors is increased while minimizing the travel time and the distance travelled using genetic algorithms. Such a system is used to compensate for coverage loss when sensors fail in the field.\\
In~\citep{li2007mesh, cui2004swarm}, a more decentralized approach for relocation of sensors is followed in order to maintain coverage of a sensor network. The sensors periodically broadcast their locations and identifiers to their neighbours and construct a Voronoi diagram. Voronoi polygons are computed using the received information. Once a node finds a hole in the Voronoi diagram, i.e. a relatively large polygon, the relocation of a sensor is initiated. In~\cite{cui2004swarm}, a simulation of an odor localization scenario with a group of mobile robots is presented. The authors focus on using fuzzy logic to decide which direction to move to in order to eventually localize the source. They assume that measurements from each agent are easily available to other agents in the swarm wherefore the agent to agent communication aspect is not adequately addressed.

In contrast to such approaches we here present a method to maximize information about the environment collected by a swarm based on a bio-inspired communication mechanism. 
The CIMAX algorithm differs from existing approaches in the following ways: 
1) the swarm has no direct access to global information -- there is no central entity knowing the positions of all sensors 
2) nor are agents able to receive instructions or be organized by a central entity. 
3) CIMAX maximizes the diversity or variance of measurements collected by the swarm as a whole and
4) our approach utilizes not only the content of received signals but also its properties.
We present both numerical simulation and robotic experiments to validate the presented method. Furthermore, this algorithm can be embedded into the ``wave oriented swarm programming paradigm'' (WOSPP)~\citep{ronnyWOSPP}, a framework for controlling swarms using the communication mechanism we briefly introduce in Section~\ref{sec:methods}.
In Section~\ref{sec:algorithm} we present the algorithm and its implementation. The computational results and theoretical analysis of the algorithm are shown in Section~\ref{sec:simulation}, including numerical simulations in the aforementioned target environment and scenario. In Section~\ref{sec:experiments} we present the experimental setup and results which are then discussed in Section~\ref{sec:discussion}.

\section{The CIMAX algorithm}
\label{sec:methods}
The CIMAX algorithm enables a swarm of individuals with limited communication abilities to make a collective decision regarding its direction of motion in order to maximize the information accessible to the collective. The fundamental communication mechanism presented here is inspired by slime mold ({\it dictyostelium discoideum}) and fireflies ({\it lampyridae}) and has previously been used to design various algorithms~\citep{varughese2016, ronnyWOSPP}. \cite{ronnyWOSPP} unified various swarm behaviours into one general framework called ''wave oriented swarm programming paradigm'' or WOSPP.

\subsection{Communication paradigm}
\label{sec:sub1methods}
\begin{figure}[thpb]
		\centering
		\subfigure[]{\includegraphics[scale=0.3]{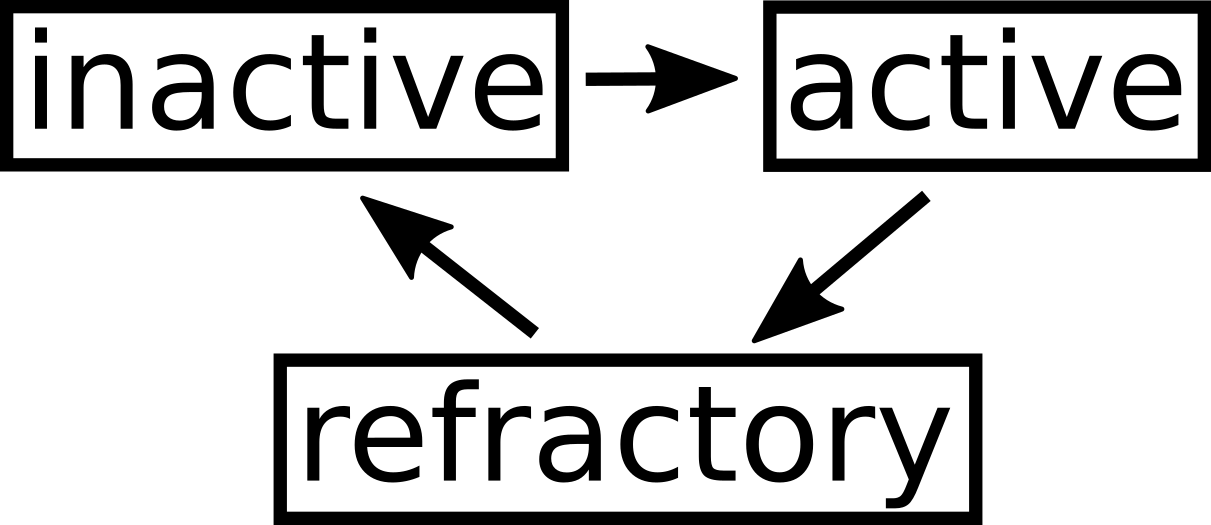}}
		\hspace{0.6cm}
        \subfigure[]{\includegraphics[scale=0.43]{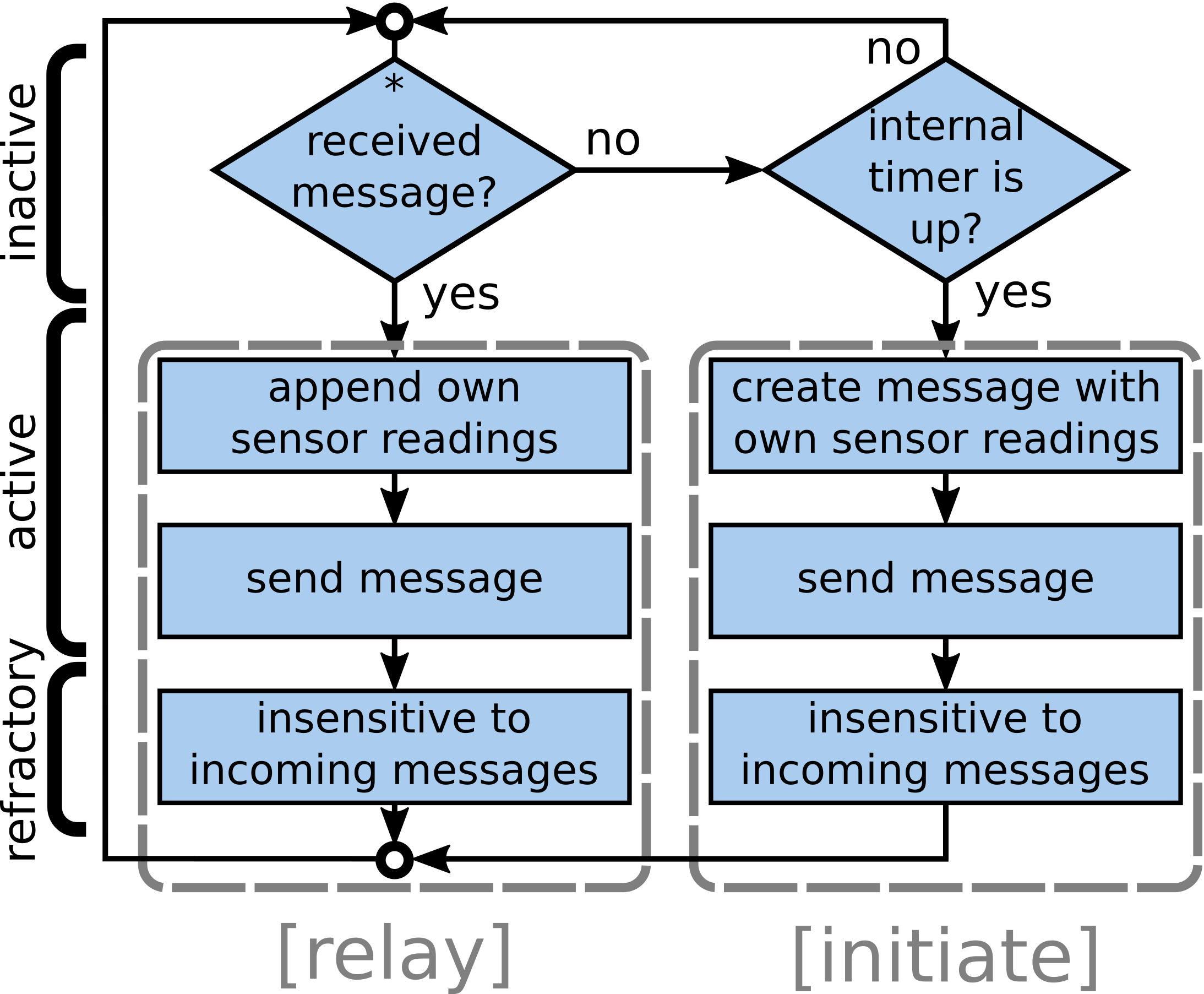}}
      \caption{The WOSPP communication mechanism. (a) Agents can be in one of the three states. From the inactive state, an incoming message or the decision to initiate a message lets an agent transition into the active state. In the active state an agent either relays an incoming message or initiates a new message. Subsequently agents enter the refractory state, being insensitive to incoming messages for a finite time until transitioning to the inactive state again. The conceptual operating structure of an agent is illustrated in (b).}
      \label{fig:state_cycle}
\end{figure}

In the WOSPP communication paradigm, all agents can enter three different states similar to the behavior of slime mold ({\it dictyostelium discoideum}): An ``inactive'' state in which agents are receptive to incoming communication, an ``active'' state where they send or relay a signal, which is followed by a ``refractory'' state where agents are temporarily insensitive to incoming signals. This communication mechanism is schematically shown in Figure~\ref{fig:state_cycle}.
Agents initiate a signal randomly by initially setting a timer within $t_p\in(0,t_{p}^{max}]$. In this manner, each agent initiates the sending of a signal at least once within a time period $t_p^{max}$ (maximum the timer can randomly be set to), which we refer to as 'cycle'.

\begin{figure}[thpb]
		\centering
		\subfigure[Time: $0 \ s$]{\includegraphics[scale=0.32]{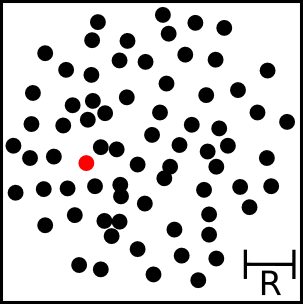}}
		\subfigure[Time: $2 \ s$]{\includegraphics[scale=0.32]{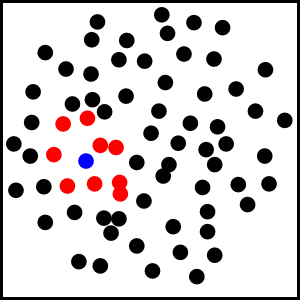}}
		\subfigure[Time: $4 \ s$]{\includegraphics[scale=0.32]{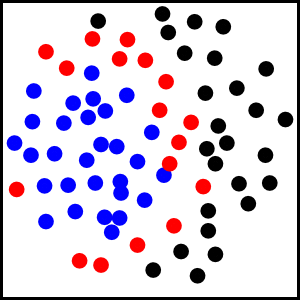}}
		\subfigure[Time: $11 \ s$]{\includegraphics[scale=0.32]{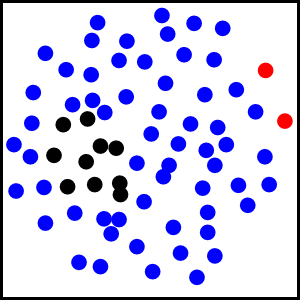}}
		\subfigure[Time: $16 \ s$]{\includegraphics[scale=0.32]{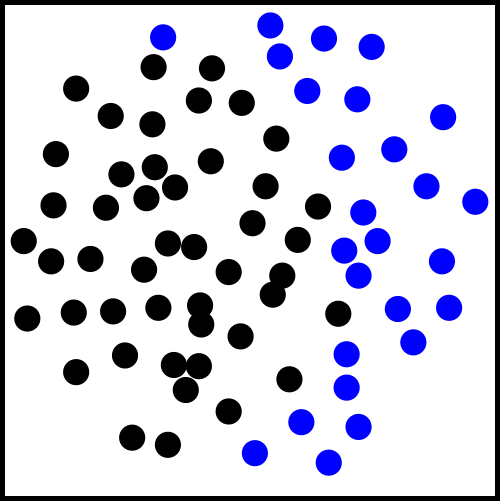}}
		\subfigure[Time: $16 \ s$]{\includegraphics[scale=0.32]{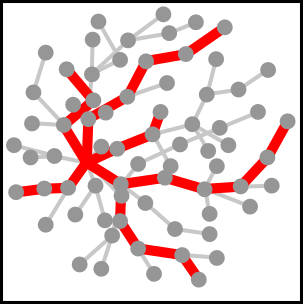}} 
      \caption{Illustration of wave based communication. In (a) almost all agents are in the inactive state, shown in black, except one agent which broadcasts a message, i.e. enters the active state, shown in red. It afterwards transitions into the refractory state, shown in blue. Neighboring agents receive the signal and switch to the active state as shown in (b) and (c). The signal spreads in a wave-like manner. In (d) the initiating agent switches from the refractory state into the inactive state again. Due to a fixed duration of the refractory state, the transition to the inactive state as well spreads in a wave like manner, shown in (d) and (e). In (f) several trajectories along which the signals were broadcast are shown as red lines.\\
      The perception range $R$ of an agent is shown as bar in the bottom right corner in (a). Times [s]: (a) 0, (b) 2, (c) 4, (d) 11, (e) 16. Parameters: number of agents $N=80$, physical size of the swarm in units perception range $R_s=5\,R$, refractory time in units timesteps $t_{ref}=10\,s$. }
      \label{fig:scrollwave}
\end{figure}

The three states of agents ultimately allow wave-like propagation of signals through the swarm as shown in Figure~\ref{fig:scrollwave}. Signals are solely received by agents in close neighborhood, i.e. within perception range $R$ of the sender and subsequently relayed thus propagating through the system. After receiving a signal agents relay it with a delay of one timestep $t_{delay}=1\,s$ which we use in the following as basic unit for time.

The refractory state assures that a signal will neither 'flood' a swarm, i.e. signals will not (re)activate the initial sender, nor periodically propagate through the swarm e.g. as a spiraling wave.  In Figure~\ref{fig:scrollwave} (a)-(e) a temporal sequence of a signal propagating through a swarm is shown. Figure~\ref{fig:scrollwave}(f) shows several trajectories signals took in the signal propagation in panels (a)-(e), indicated by red lines. For this algorithm agents need to be able to communicate with nearest neighbors, move or have some means of transportation, and have a common sense of direction.

\subsection{The Algorithm}
\label{sec:algorithm}

In the following we present the algorithm for maximizing the information accessible to, or collected by the swarm. For this scenario we define information as diversity of measurements throughout the swarm, quantified using the variance of measurements. Thus the swarm ultimately detects diverse domains or transition areas between homogeneous domains, while uniform domains are considered redundant and providing less information.

Each agent in the swarm measures the same single quantity $X$ which we use as a generic placeholder for any environmental parameter or quantity measured by swarms. When one agent initiates a message, it sends its own measurement value as message. Neighboring agents receive the message, append their own measurements and relay the message. This way a message propagating through the system incrementally grows in length with every relay. 
With this in mind, for easier illustration of the algorithm we divide the entire procedure into three parts: ``information gathering'', ``evaluation'' and ``collective decision''. However, for implementation there are various possibilities, depending on the abilities and specific tasks of the target medium, without need of dividing.

\begin{figure*}[thpb]
		\centering
		\subfigure[Information gathering]{\includegraphics[scale=0.3]{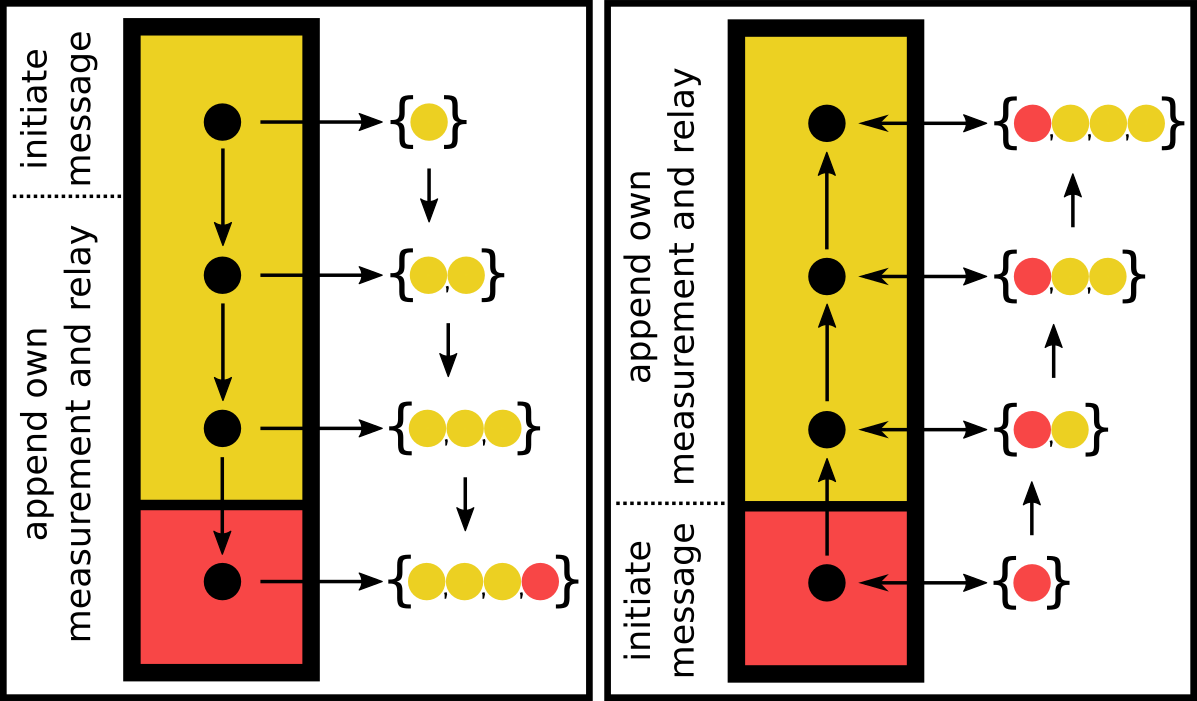}} \hfill
		\subfigure[Evaluation]{\includegraphics[scale=0.3]{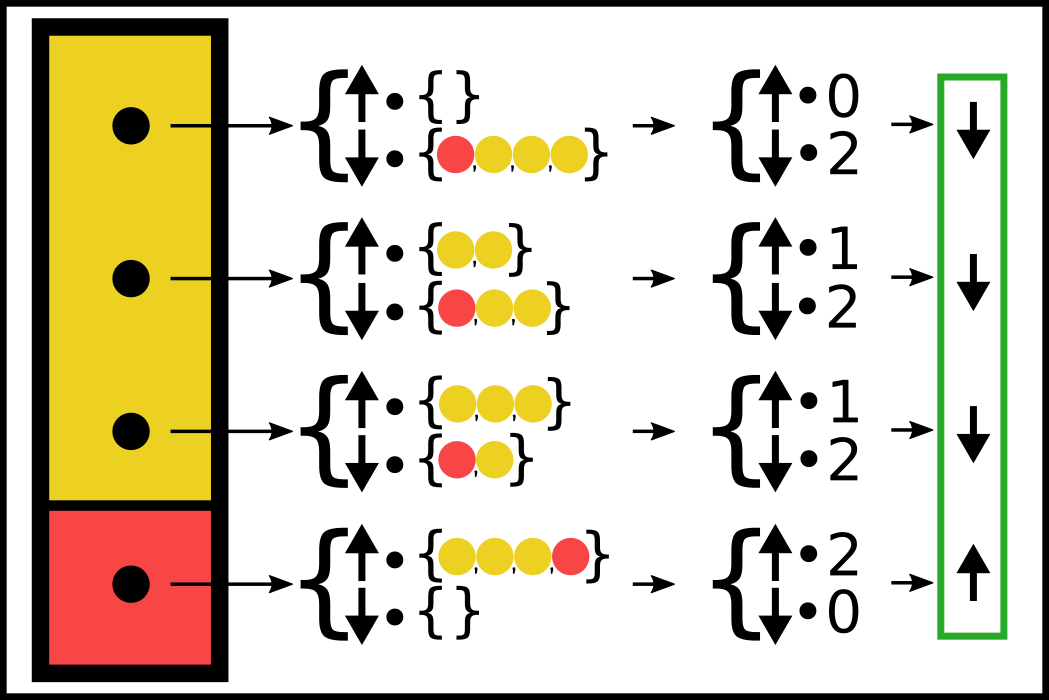}} \hfill
		\subfigure[Collective decision]{\includegraphics[scale=0.3]{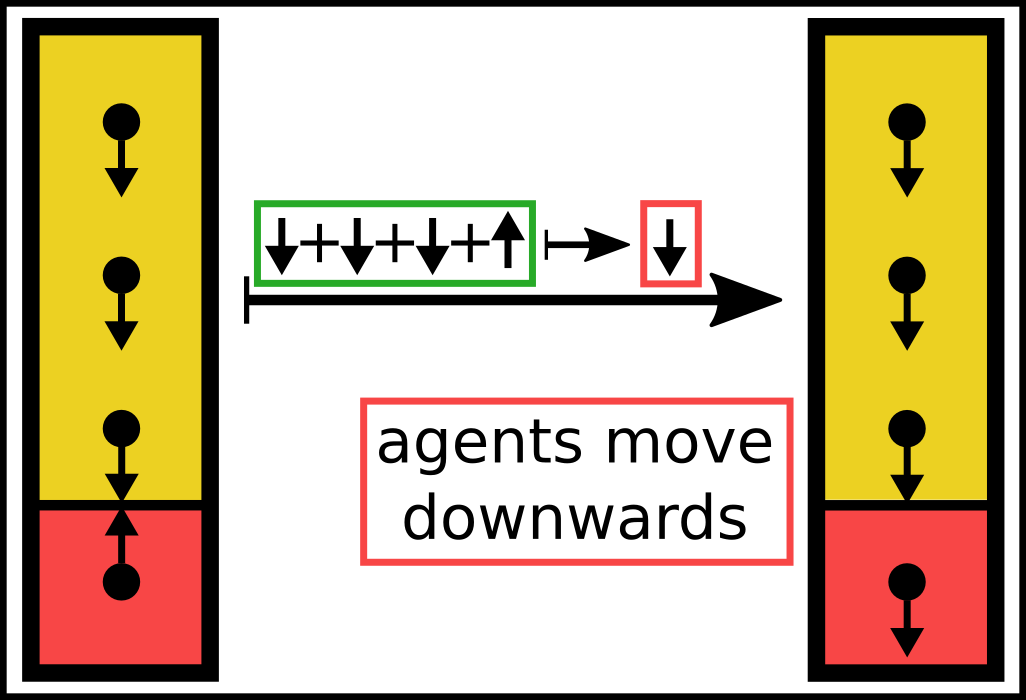}}
		\caption{The three sub-parts the algorithm can be divided into. Four agents illustrated as black dots constitute a swarm in a system with two domains, yellow and red. The colors represent two different measurements of quantity X. (a) The dispersion of two independent messages. On the left hand side the top agent (in the yellow domain) initiates a message with its own measurement, illustrated as a yellow dot in curly brackets next to the agent. The message propagates from agent to agent, each of which appends its own measurement (here depicted as color). On the right hand side the same scenario is shown only this time the bottom agent, in the red domain, initiates the message which then propagates upwards. (b) The evaluation of the two messages. The diversity of a message is illustrated as the number of different measured colors. The top agent received no messages from upward direction and thus considers a weight of $w=0$ colors for upward, however a weight of $w=2$ colors for downward. Its preferred direction therefore is down which is indicated by the arrow in the green box on the right hand side. All agents calculate their preferred directions in this way. (c) All agents communicate their preferred direction (not explicitly shown) and ultimately agree on a direction to move. Since three agents prefer to move downwards and one agent prefers upwards, the resulting common direction is downwards.}
      \label{fig:algorithm}
\end{figure*}

The three parts of the algorithm are exemplary illustrated in Figure~\ref{fig:algorithm} for a swarm of $N=4$ agents. The agents, represented by black circles, are arranged in a line. They are able to measure a quantity $X$ of the environment which is represented by the background colors red and yellow.

Four agents, represented by black circles, constitute a one-dimensional swarm within a system with two domains, yellow and red, representing two different measurements X.

\begin{figure}[thpb]
		\centering
		\subfigure[]{\includegraphics[scale=0.39]{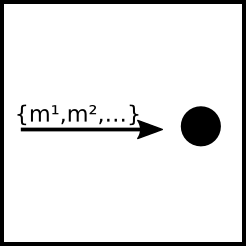}}
		\subfigure[]{\includegraphics[scale=0.39]{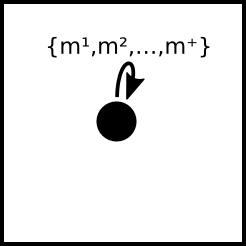}}
		\subfigure[]{\includegraphics[scale=0.39]{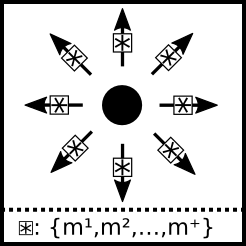}}
		\caption{(a) An agent (black circle) receives an incoming message with measurements $m^i$. (b) It then appends its own measurement and stores the message before in (c) it broadcasts this extended message.}
      \label{fig:message}
\end{figure}

\begin{itemize}
    \item \textbf{Information gathering}: agents randomly (in time) initiate sending a message containing their own sensor readings. Each agent which has received this message stores the received information as well as the direction from which it received the message. Finally, each agent appends its own sensor readings before then broadcasting it to its neighbours. This process is schematically shown in Figure~\ref{fig:message}, resulting in a dispersion of information about the sensor readings of agents throughout the whole swarm.
    
    \item \textbf{Evaluation}: agents evaluate the stored messages with respect to the directions from which they were received. 
    Agents then determine the diversity of the content of all messages associated with a certain direction. Depending on the systems characteristics this is practically done e.g. by calculating the variance of all elements contained by those messages. 
    The calculated diversities serve as ``weights'' for all directions. Agents finally consider the direction with largest weight (e.g. variance) as their preferred direction to move towards.
    Figure~\ref{fig:algorithm} (b) shows the evaluation of the two messages initiated in~\ref{fig:algorithm} (a).

    \item \textbf{Collective decision}: agents agree on a common direction to move towards a target location, based on the individual preferences of directions. One option is to let agents communicate their opinions on a preferred direction to the neighbors. Those then, instead of relaying a message, simply change their own preferred direction by a small factor towards the received direction. This way opinions 'diffuse' through the swarm letting it converge to a common opinion. Figure~\ref{fig:algorithm} (c) shows the result of a collective decision on the example shown in~\ref{fig:algorithm} (a) and (b).

\end{itemize}

Algs. \ref{alg:info_gathering}, \ref{alg:evaluation} and \ref{alg:coll_decision} show the pseudo-code for the three parts ``information gathering'', ``evaluation'' and ``collective decision'', respectively. Please note that the presented pseudo-code is an exemplary implementation of the algorithm and does not exclude alternative ways of implementing it.

\begin{algorithm}[tbph]
		  Mode $\leftarrow$ ``information gathering''\;
		  state $\leftarrow$ $inactive$\;
		  timer($t_{p}$) $\leftarrow$ random integer $\in (0,\ t_{p}^{max}]$\; 
		 \While{Modus = ``information gathering''}
		 {
		  decrement timer($t_{p}$)\;
		  \If{agent in refractory state}
		  {
		   wait for refractory\_time\;
		   \If{refractory\_time is over}{
		   state $\leftarrow$ inactive
		   }
		 }
		  \If{agent in active state}
		  {broadcast message\;
		  state $\leftarrow$ refractory}
		  \If{agent in inactive state}{listen for incoming pings\;
		  \If{message received}{
		   state $\leftarrow$ active\;
		   append own measurements to received message $i$\;
		   calculate variance $V_i$ in measurements contained by message $i$\;
		   store variance $V_i$ with respect to the direction of reception of the message, $V_{i}^{dir}$\;
		   }
		   }
		  \If{timer($t_{p}$) $\leq$ 0}
		  {state $\leftarrow$ active\;
		  create empty message and append own measurements to message\; 
		  }
		 }
		 \caption{Pseudo-code of ``information gathering''}
		 \label{alg:info_gathering}
\end{algorithm}

\begin{algorithm}[tbph]
   Mode $\leftarrow$ ``evaluation''\;
   calculate average of variances $\overline{V^{dir}}$ of for each direction of reception $dir$\;
   preferred direction $dir_{preferred}$ $\leftarrow$ choose direction associated with largest average variance $\overline{V^{dir\_{preferred}}}={\textrm{max}}\{\overline{V^{dir}}\}$\;
   empty storage\;
   \caption{Pseudo-code for ``evaluation''}
   \label{alg:evaluation}
\end{algorithm}

\begin{algorithm}[tbph]
		  Mode $\leftarrow$ ``collective decision''\;
		  state $\leftarrow$ $inactive$\;
		  timer($t_{p}$) $\leftarrow$ random integer $\in (0,\ t_{p}^{max}]$\; 
		 \While{Modus = ``collective decision''}
		 {
		  decrement timer($t_{p}$)\;
		  \If{agent in refractory state}
		  {
		   wait for refractory\_time\;
		   \If{refractory\_time is over}{
		   state $\leftarrow$ inactive
		   }
		 }
		  \If{agent in active state}
		  {broadcast preferred direction $dir_{preferred}$\;
		  state $\leftarrow$ refractory}
		  \If{agent in inactive state}{listen for incoming pings\;
		  \If{message received}{
		   state $\leftarrow$ active\;
		   adjust own preferred direction by 10 \% towards preferred direction contained by the incoming message;
		   }
		   }
		  \If{timer($t_{p}$) $\leq$ 0}
		  {state $\leftarrow$ active\;
		  }
		 }
		 \caption{Pseudo-code of ``collective decision''}
		 \label{alg:coll_decision}
\end{algorithm}

\section{Simulation}
\label{sec:simulation}

In this section we first present the behavior of a swarm in systems consisting of a discrete and a smooth linear transition, respectively, to give an intuitive understanding of its behavioral dynamics. We then examine a computational scenario close to a real application case.\\
We consider a swarm of N=61 agents within a 2-dimensional space. Each agent has a perception range of $R$. Agents in the swarm are distributed in a circular area of diameter $D=6R$. 
We chose the number of agents in the swarm $N$ relative to $D$ in a manner such that agents on average have five neighbors within perception range in order to ensure sufficient connectivity within the swarm. 
Every negotiation period, after the swarm decided for a direction to move, the swarm moves by a step of length $s=0.33 R$ along this direction. For simplicity we let the swarm move as a whole without changing the agents' relative positions. Hence we exclude any interaction between agents other than communication and treat agents as point particles. Each agent is able to measure a dimensionless quantity $X$ in the system which we use as placeholder for any environmental parameter or quantity. 
Finally, in the following we quantify diversity using the messages stored by agents. We define diversity $V_k$ associated with an agent $k$ as the variance of the measurements $m_j^k$ contained by all stored messages of this agent
\begin{align}
    V_k = \frac{1}{n}\sum_{j=1}^n (m_j^k - \overline{m^k})^2 \;\;\;,
\end{align}
where $n$ represents the total number of measurements $m_j^k$ and $\overline{m^k}$ represents the average of those measurements $\overline{m^k} = \frac{1}{n}\sum_{j=1}^n m_j^k$.

\subsection{Discrete distribution of environmental factors}
\label{sec:simulation_sharp}
\begin{figure}[thpb]
		\centering
		\subfigure[]{\includegraphics[scale=0.6]{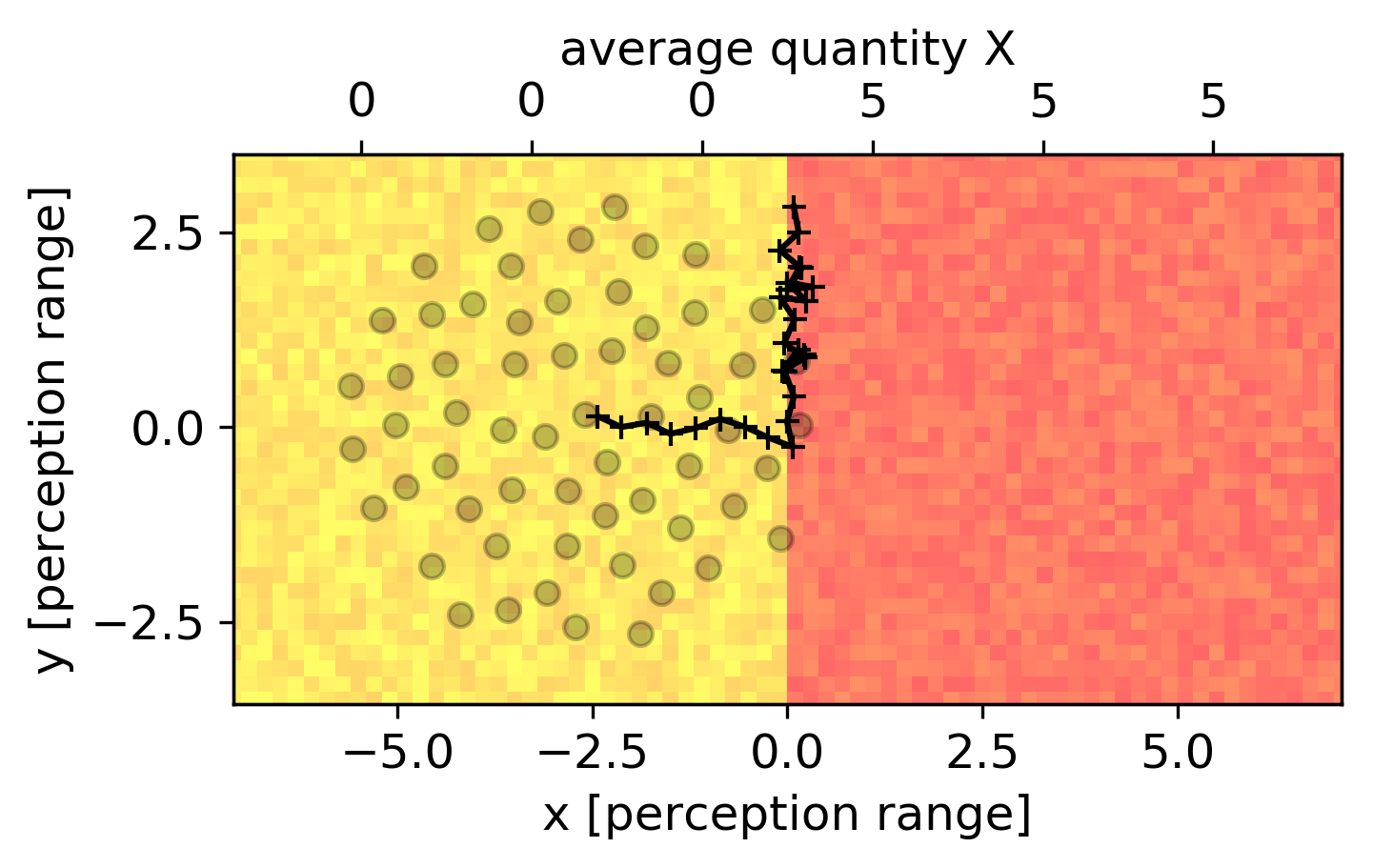}}
		\subfigure[]{\includegraphics[scale=0.6]{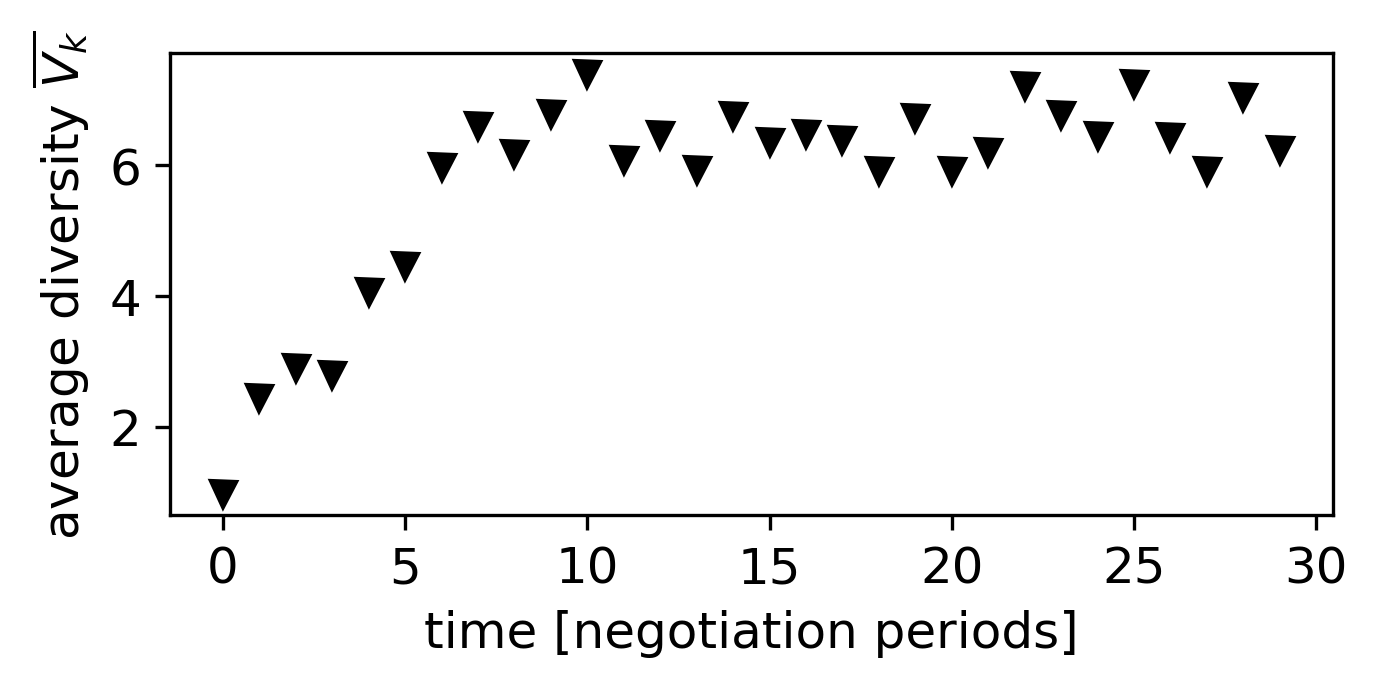}}
		\caption{(a) shows the trajectory of the average position of the swarm (+) within a system with two domains, yellow with $X\in[-0.5,0.5]$ and red with $X\in[4.5,5.5]$ Initially the swarm moves towards right until it reaches the border between the domains. The remaining simulation time it moves randomly along this border. In light grey circles the swarm at its initial position (with center of the swarm at $x\approx-2.5$)
		is shown, each circle representing one agent. [$N=61$, negotiation period: 2 cycle lengths]. (b) shows the diversity averaged over all agents in the swarm $\overline{V_k}$ against time $t$ from an observer’s perspective. Initially the diversity is small ($t=0$, $\overline{V_k}=0.2$) as most agents have similar measurements in the yellow domain with quantity $X$ levels fluctuating around 0. With increasing time, more agents have different measurements as the swarm approaches the border.}
      \label{fig:discrete_transition}
\end{figure}

In Figure~\ref{fig:discrete_transition} the scenario of a swarm close to a sharp transition of a measured quantity $X$ is shown, left hand side in yellow for low values, right hand side in red for high values of $X$. Everywhere the quantity $X$ is subject to a small time dependent random noise $-0.5<\xi(t)<0.5$. The center of mass of the swarm, represented by a black ``$+$'', is initially at position $(x,y)=(-2.5,0)$, in the yellow domain. 
All agents are illustrated as grey dots at their initial position (with center of mass of the swarm at $(x,y)=(-2.5,0)$).
In the following we use ``the position of the center of mass of the swarm'' synonym with ``the position of the swarm''.

In the beginning of the simulation the swarm moves straight to the right, towards the border of the two domains. From there at $(x,y)\approx(0,0)$ it moves upwards along the border of the two domains in a less directed manner, effectively performing a one-dimensional random walk.
Figure~\ref{fig:discrete_transition} (b) shows the average diversity $\overline{V_k}$ within the swarm (as viewed by an external and all-knowing observer) against time. Initially the average diversity is close to $\overline{V_k}=0$ and increases until $t=8$ where it reaches a plateau around $\overline{V_k}=4.5$. This corresponds to the point when the swarm reached the border. Please note that the swarm is not attracted by domains of higher values of $X$, but instead by largest average diversity of measurements and therefore moves towards the transition.

\begin{figure}[thpb]
		\centering
		\subfigure{\includegraphics[scale=0.56]{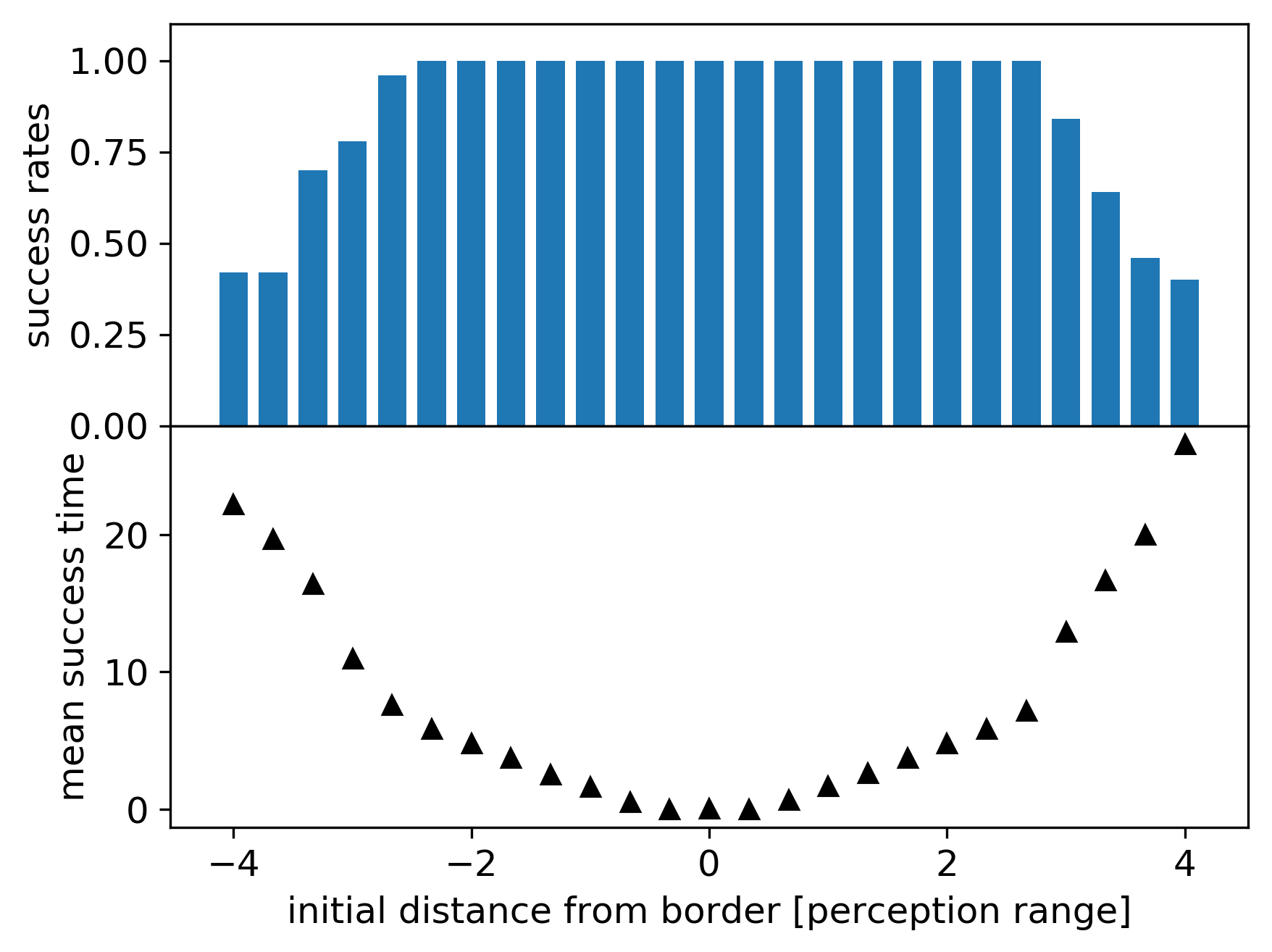}}
		\caption{The top graph shows the success rates of the swarm moving the border between the two domains vs. its initial position relative to the border in our simulation experiment depicted in Fig. \ref{fig:discrete_transition}. A simulation is counted as successful if the swarm reaches a distance from the border smaller than $|x|<0.33\,R$. For initial distances from the border smaller than $|x_{init}| < 2.5$, the swarm consistently succeeds in finding the border. After 50 negotiation periods (corresponding to 100 cycles) a simulation was stopped and counted as failed. The bottom graph shows the mean time for a swarm until it reaches the border. Each data is the result of 50 independent simulations.}
      \label{fig:discrete_transition_analysis}
\end{figure}

In Figure~\ref{fig:discrete_transition_analysis} we show the rate in which a swarm successfully reaches the border between the two domains. We count a simulation as successful if the center of mass of the swarm reaches a distance to the border smaller than $|x| < 0.33R$ within a finite simulation time of $t_{fin}=100\,t_p^{max}$, i.e. the time in which the swarm can take 50 steps.
In Figure~\ref{fig:discrete_transition_analysis} the success rates (histogram in top figure) and corresponding mean time until success (bottom figure) of a simulation is shown for different initial distances of the swarm from the border. 
For initial distances smaller than $|x_{init}| < 2.5\,R$ the success rates are 1 and the corresponding success time decreases linearly. This shows, that if the swarm initially perceives the other domain (yellow and red domain as shown in Fig.~\ref{fig:discrete_transition}, respectively), it consistently moves there directly. For distances further away the swarm randomly moves around and by chance perceives the respective other domain.

\begin{figure}[thpb]
		\centering
		\subfigure{\includegraphics[scale=0.8]{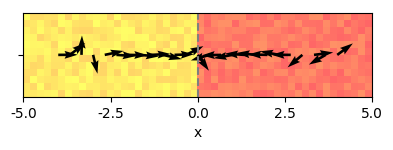}}
		\caption{The preferred direction of a swarm in a system with two domains, yellow ($X \in [-0.5,0.5]$) and red ($X \in [4.5,5.5]$). Every arrow indicates the preferred direction of a swarm with its center at the arrow’s position. The arrows were calculated each with a single simulation with negotiation periods of 4 cycle lengths. For $|x|\gtrsim2.5$ the swarm moves randomly, for $|x|\lesssim2.5$ it can perceive the other domain and moves towards it - from both direction respectively towards the border at $x=0$.}
      \label{fig:discrete_transition_vectorfield}
\end{figure}

This consistent behavior allows us to illustrate the expected behavior of a swarm close to the border as shown in Figure~\ref{fig:discrete_transition_vectorfield}. It shows the preferred direction of such a swarm as arrows. Each arrow represents the preferred direction of a swarm with its center at the arrow's location. For distances from the border $|d|>2.5$ the preferred direction is random since no agent in the swarm is located in the respective other domain and thus the swarm has no information about its existence. In this case the swarm is located in an almost uniform area and thus does not develop a preferred direction. For distances from the border $|d|<2.5$, the swarm moves towards the border.

\subsection{Gradient distribution of environmental factors}
\label{sec:simulation_gradient}

\begin{figure}[thpb]
		\centering
		\subfigure[]{\includegraphics[scale=0.6]{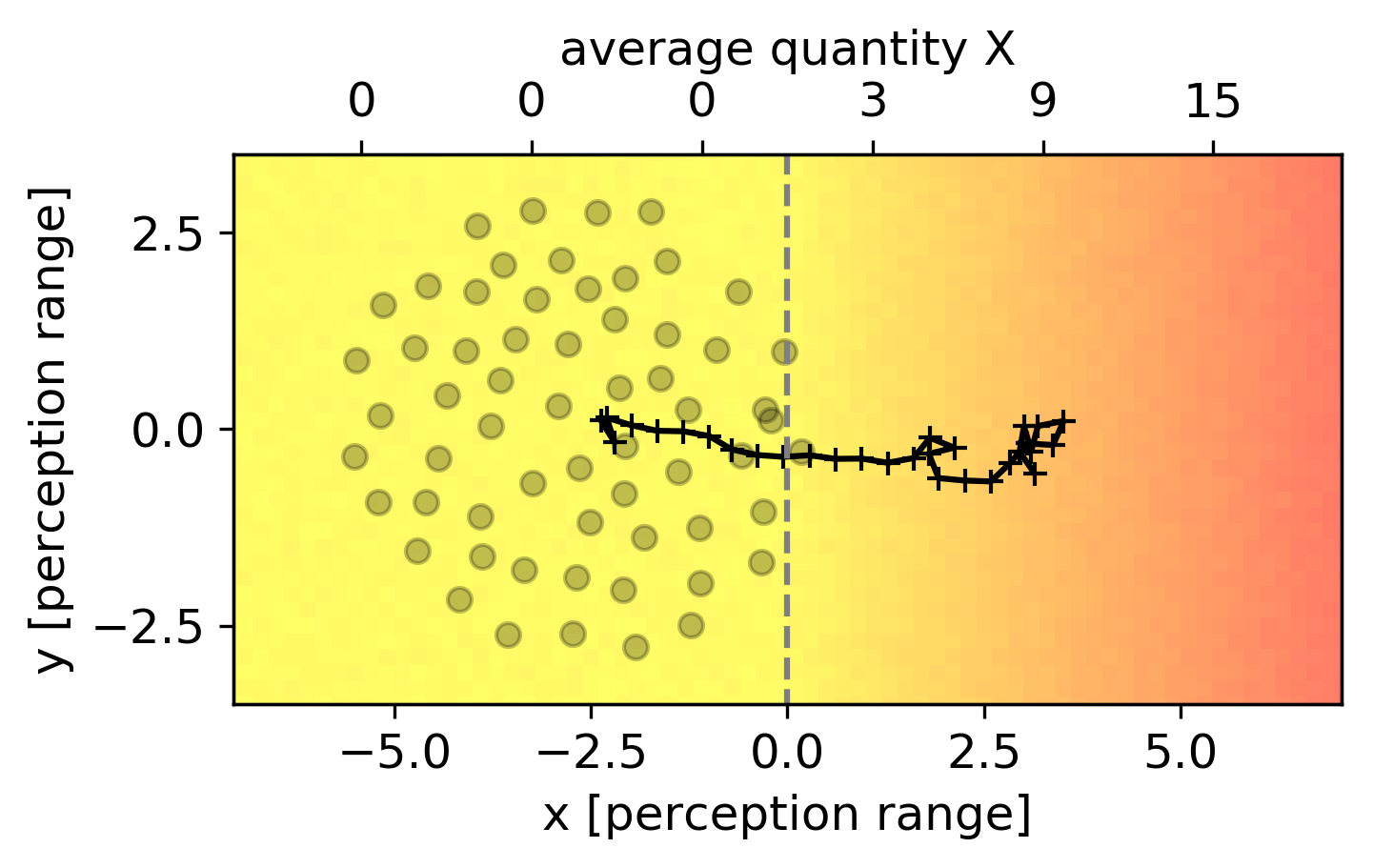}}
		\subfigure[]{\includegraphics[scale=0.6]{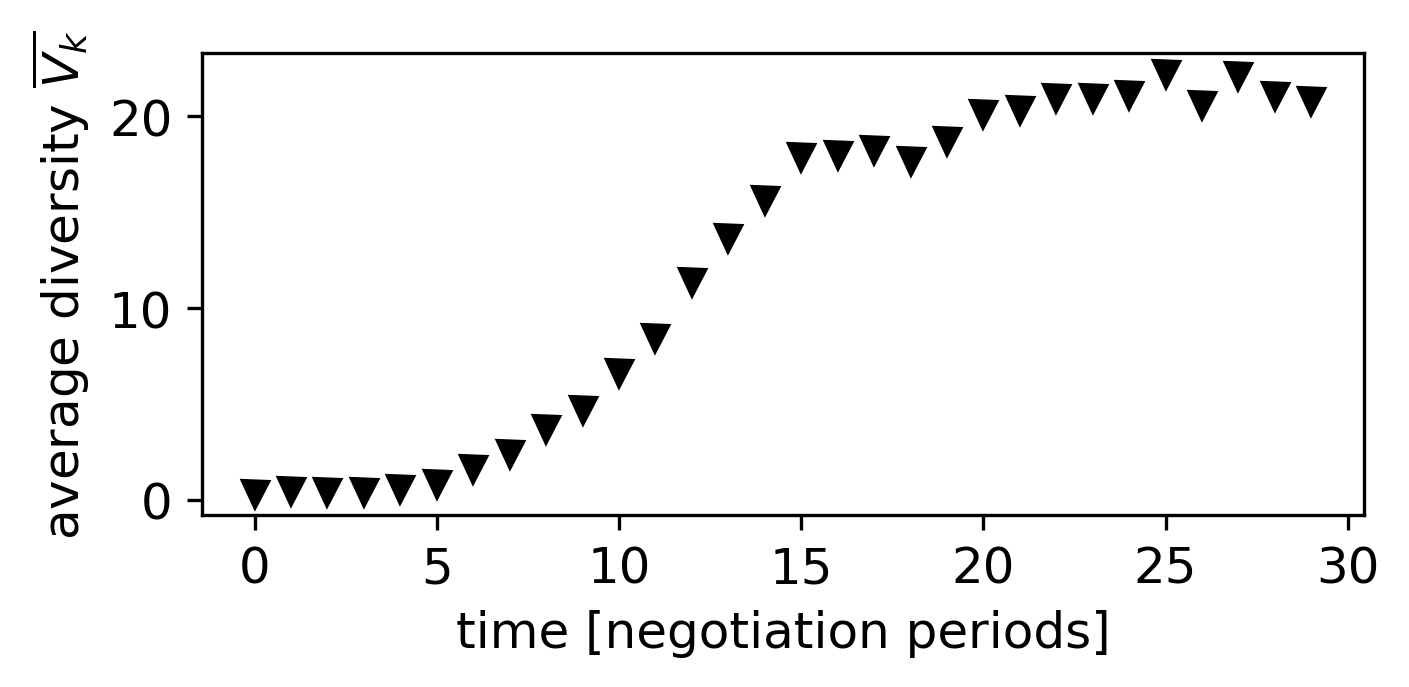}}
		\caption{(a) The trajectory of the center of mass of the swarm (+) in a system with two domains, for $x\leq 0$ the environment exhibits values fluctuating between $X\in[-0.5,0.5]$, for $x>0$ the average values in $X$ linearly increase with increasing $x$. Initially the swarm moves towards the right until at $x\approx2.5$ it moves randomly. The dashed line indicates the border between the area of on average uniform levels of $X$ (left) and the (along the x-axis) linearly increasing domain. For the initial position of the swarm with center of mass at $(x,y)\approx(-2.5,0)$ the agents of the swarm are shown as light grey circles. [$N=61$, negotiation period: 2 cycle lengths] (b) shows the average of the agents' diversities $\overline{V_k}$ over time. Initially $\overline{V_k}$ is close to zero as the swarm is located in an almost uniform area. With increasing time as the swarm moves towards the right, the average diversity increases until around $t=15$ it saturates and fluctuates around $\overline{V_k}=8$. This corresponds to the random walk of the swarm.}
      \label{fig:smooth_transition}
\end{figure}

In Figure~\ref{fig:smooth_transition} a swarm close to a gradient in $X$ is shown. For $x<0$ the system exhibits fluctuating values in $X\in[-0.5,0.5]$ , for $x\geq0$ the temporal average in $X$ linearly increases. The swarm initially starts at position $(x,y)=(-2.5,0)$ and moves towards the right in a directed manner. For $x\gtrsim2.33$ the swarm moves less directed and effectively performs a random walk. 
In Figure~\ref{fig:smooth_transition} (b) the diversity averaged over all agents in the swarm $\overline{V_k}$ is shown against time. It increases from $\overline{V_k}\approx0.3$ until at $t=15$ (when it starts moving randomly) it reaches a plateau where it fluctuates around $\overline{V_k}=20$.

\begin{figure}[thpb]
		\centering
		\subfigure{\includegraphics[scale=0.56]{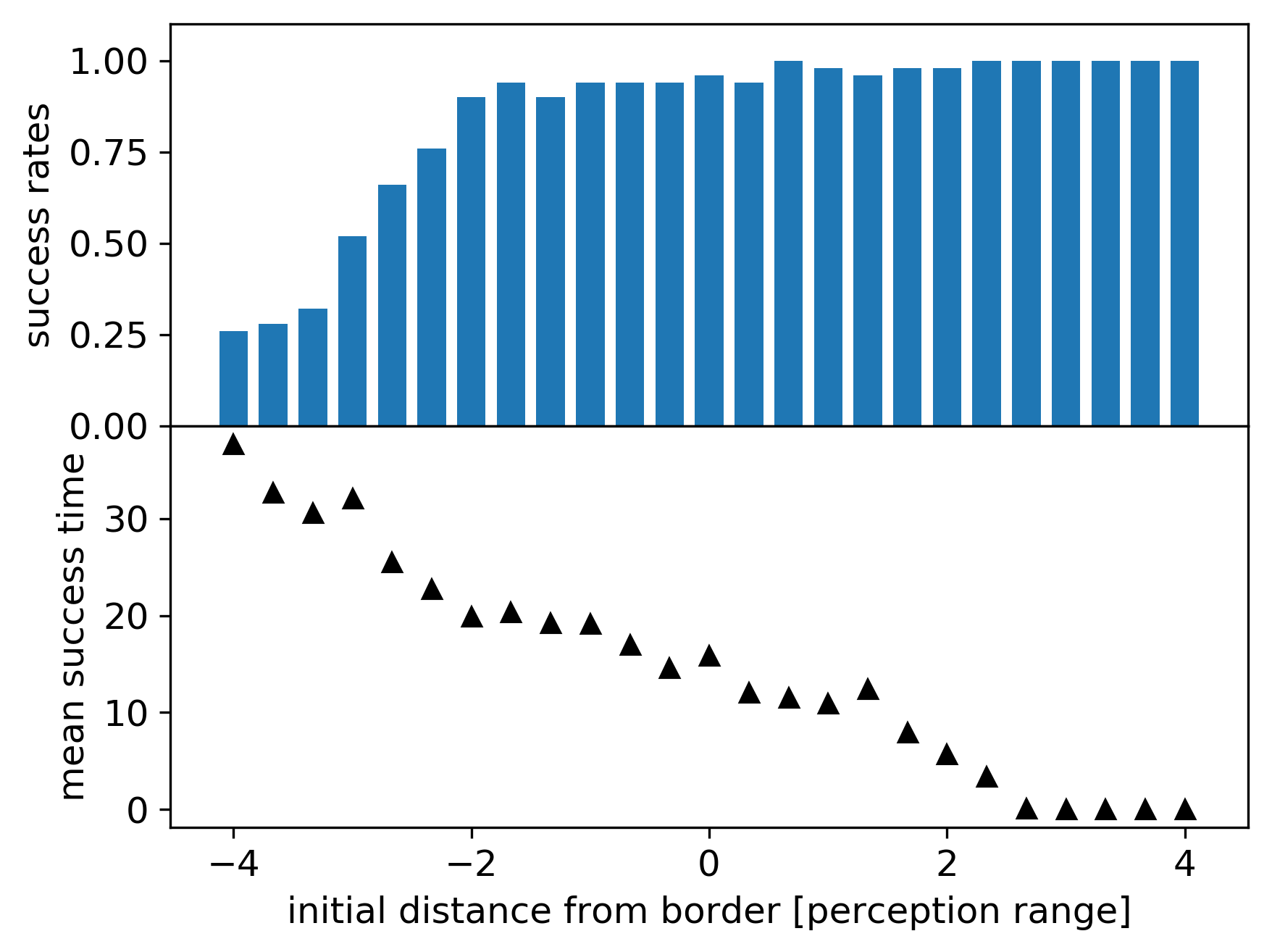}}
		\caption{The top graph shows the success rates of the swarm reaching the border (as depicted in Fig.~\ref{fig:discrete_transition}) between the two domains versus the initial position of the swarm. For each distance we conducted 50 independent simulations. As success we counted simulations in which the swarm reached a position $x\geq2.33$, corresponding to an average diversity of $\overline{V_k}\approx 8$ (as shown in Figure~\ref{fig:smooth_transition}). For $x\gtrsim2.33$ the swarms performs a random walk. After 50 negotiation periods (corresponding to 100 cycles) a simulation was stopped and counted as failed. The bottom graph shows the mean time for a swarm until it succeeds, only taking into account successful simulations.}
      \label{fig:smooth_transition_analysis}
\end{figure}

In Figure~\ref{fig:smooth_transition_analysis} we show the rate in which a swarm successfully maximizes its average diversity $\overline{V_k}$. We count a simulation as successful if the swarm reaches a position of $x \geq 2.33$ within a finite simulation time of $t_{fin}=100\,t_p^{max}$, i.e. the time in which the swarm takes 50 steps. 
In Figure~\ref{fig:smooth_transition_analysis} the success rates (top histogram) and corresponding mean time until success (bottom graph) of a simulation is shown for different initial distances of the swarm from the onset point of the linear-increase domain. For initial positions $x_{init}\geq 2.33$ the swarm succeeds instantly as its initial position already fulfills the condition for success. For $-2\lesssim x_{init} < 2.33$ the swarm succeeds in the majority of conducted simulations, the success times increase with decreasing distance from the border between the two domains. At $x_{init}\lesssim -2$ for decreasing distance from the border the success rates decrease significantly, at $x_{init}=-4$ they reach $25\%$. For $x_{init}\lesssim-2$ the swarm is too far away from the domain of increasing values in $X$ and therefore does not perceive it anymore. By chance it moves closer to the domain of increasing values in $X$ and ultimately succeeds, i.e. reaches $x\geq2.33$ within simulation time. Only successful simulations were considered when calculating the mean success times.

\begin{figure}[thpb]
		\centering
		\subfigure{\includegraphics[scale=0.8]{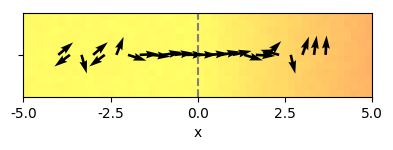}}
		\caption{The preferred direction of a swarm in a system with (towards the right hand side) linearly increasing average levels of $X$. For $x<0$, $X$ fluctuates around $X\in[-0.5,0.5]$. For $x\geq0$ the average levels in $X$ increase linearly with increasing $x$. The dashed line indicates the onset of the increase. Every arrow indicates the preferred direction of a swarm with its center at the arrow’s position. Each arrow was calculated with a single simulation with a negotiation period of $4$ cycle lengths. For $x\lesssim -2$ the swarm moves randomly as it does not perceive the gradient domain. For $-2\gtrsim x\gtrsim 2$ the swarm is attracted to the gradient domain and moves towards the right, effectively maximizing its average diversity. For $x\gtrsim 2$ the swarm moves randomly. At this point, both directions (along the x-axis) exhibit the same linear gradient. Since the swarm detects the variance in measurements instead of absolute values, both directions are equivalent.}
      \label{fig:smooth_transition_vectorfield}
\end{figure}

Instead of diffusing along the border as it is the case for a sharp transition, for this gradient the swarm diffuses in both dimensions given that $x\geq2.3$. As soon as the swarm is entirely on a linear gradient (for $x>2.3)$, both directions (to lower and to higher values, respectively) provide the same average diversity. This is implicitly shown in Figure~\ref{fig:smooth_transition_vectorfield} where each arrow denotes the preferred direction of a swarm with center at its position. For $x<-2.2$ the swarm moves randomly as it does not perceive the linear gradient (starting at the dashed line). For $x>2.2$ the swarm moves towards the right up to $x=2.2$ where it moves randomly.

\subsection{2-dimensional cloud}
\label{sec:simulation_cloud}

\begin{figure}[thpb]
		\centering
		\subfigure{\includegraphics[scale=0.7]{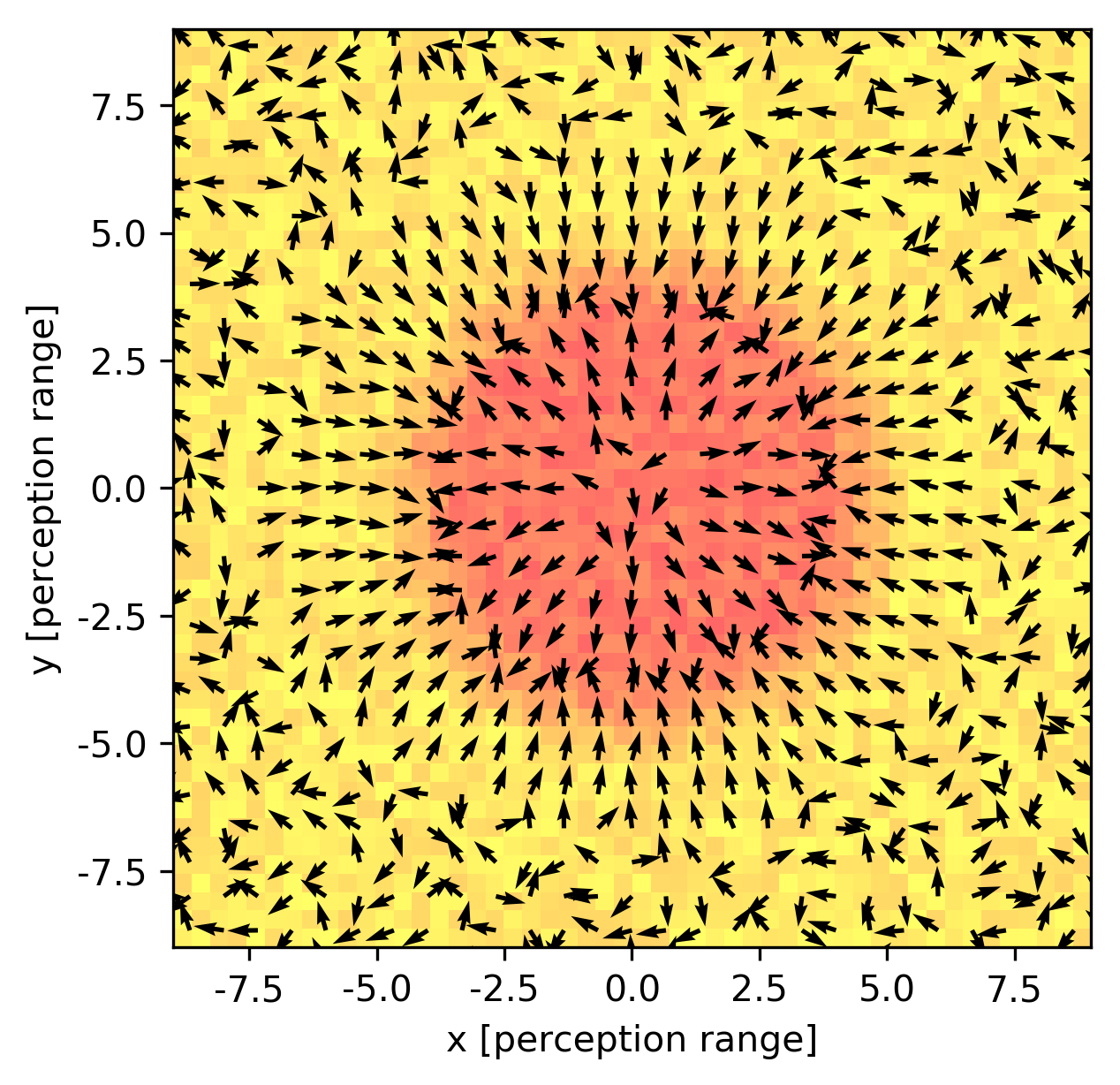}}
		\caption{Preferred direction of a swarm within a system with a circular domain of deviating levels of $X$. Each arrow represents the preferred direction of a swarm with its center it its position. Each arrow was calculated by a single simulation with negotiation period of 2 cycle lengths. The circular domain extends radially with a radius of $r=5$. The levels of $X$ linearly decrease from a maximum of 5 down to 0, to every position in the system is added a noise between 0 and 1. For distances form the center of the circular area larger than $d\gtrsim 7.5$ the swarm is too far from it to perceive it and thus does not find a coherent preferred direction, i.e. the swarm moves randomly. For $d\lesssim7.5$ the swarm consistently moves towards the border between the two domains.}
      \label{fig:cloud_vectorfield}
\end{figure}

We consider an area of radially varying values of $X$. 
In Figure~\ref{fig:cloud_vectorfield} the quantity $X$ fluctuates around a constant value $X\in[4.5,5.5]$ in the yellow domain. For radial distances from the center of the cloud between $d\in[3,4.5]$ the quantity $X$ linearly decreases to $X\in[-0.5,0.5]$. For $d<3$ the red domain exhibits $X\in[-0.5,0.5]$. 
Figure~\ref{fig:cloud_vectorfield} shows the preferred direction of a swarm as arrows, the position of each arrow indicating the center of mass of the swarm. For radial distances from the center of the cloud of $d\gtrsim7$ the swarm moves randomly as it does not detect the circular domain. For $7\gtrsim d\gtrsim 4$ the swarm moves towards the circular domain, whereas for $d\lesssim 4$ it moves radially away from it towards its border.
As soon as the swarm detects the circular domain of deviating levels, it proceeds to move towards its border where it measures the largest average diversity.

\section{Robotic Experiments}
\label{sec:experiments}

For experimental validation of the CIMAX algorithm we used aMussel robots, developed in the project subCULTron~\citep{donati2017amussels}. They communicate via modulated light and are used i.a. for examining the anoxic waters phenomenon in the lagoon of Venice~\citep{Anoxia} by diving down to the floor of the lagoon. They are equipped with a variety of sensors and communication devices~\citep{donati2017amussels}, including sensors for oxygen levels as well as ambient light. aMussels can only dive up to the water surface and down to the floor of a water body and have no other means of transportation of their own. In the field they are transported by a different type of robot which constitutes a part of the heterogeneous robotic swarm within project subCULTron~\citep{thenius2016subcultron}. \\
In the experiments presented in the following we use the robots exclusively for validating the decision making process of the CIMAX algorithm.

\subsection{Experimental setup}

\begin{figure}[thpb]
		\centering
		\subfigure[]{\includegraphics[width=7cm,height=3cm]{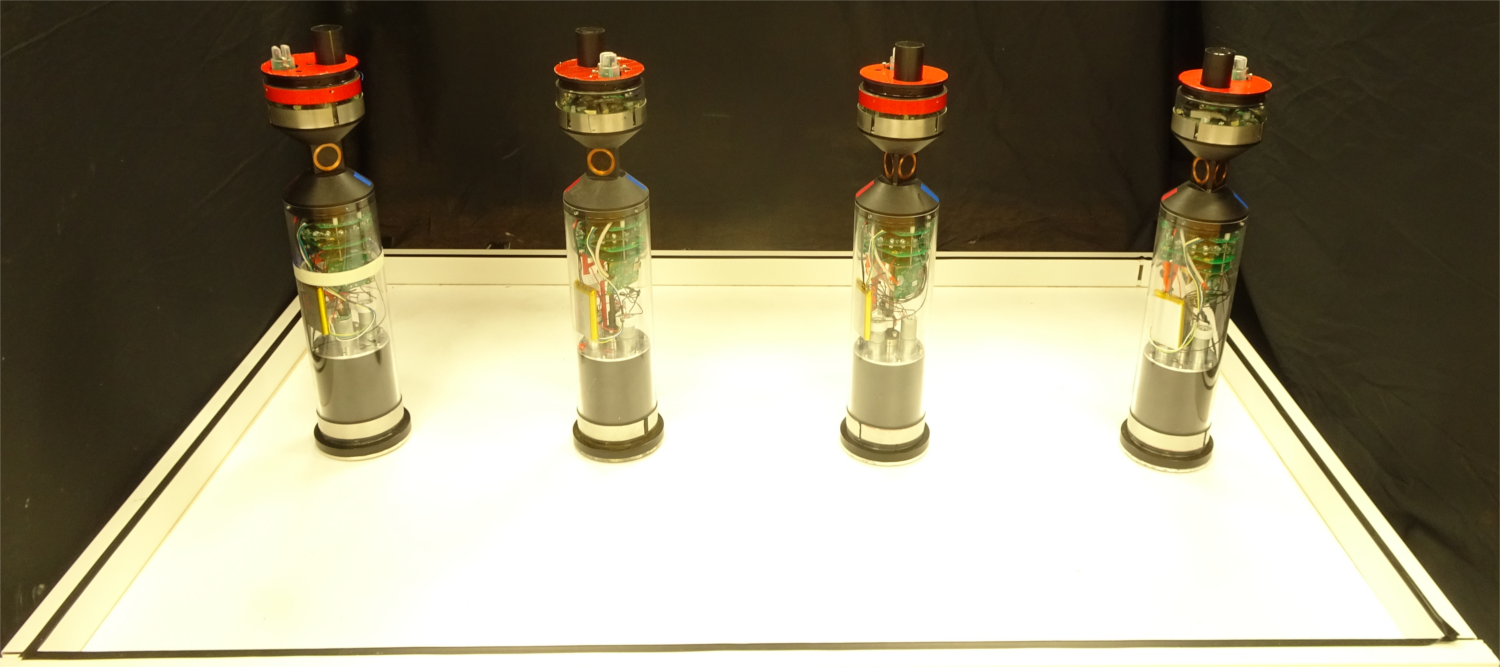}}
		\subfigure[]{\includegraphics[width=7cm,height=3cm]{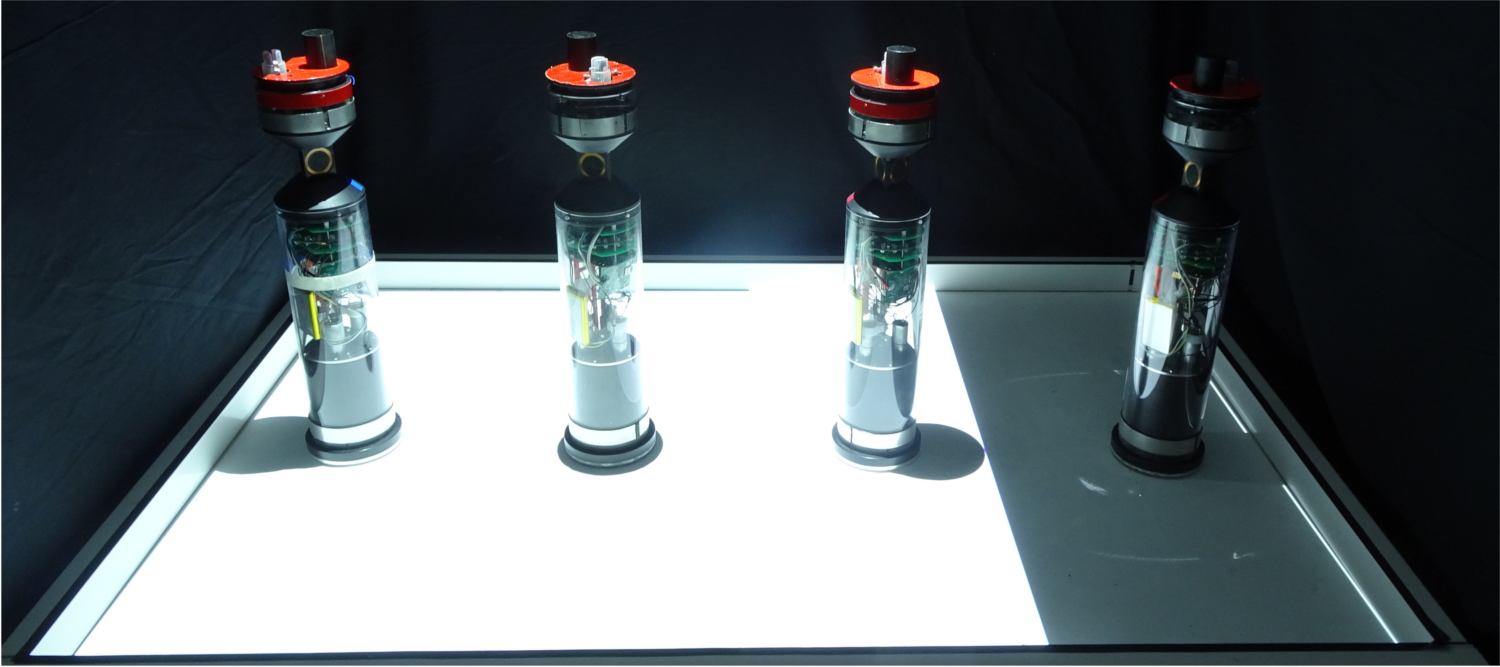}}
		\subfigure[]{\includegraphics[width=7cm,height=3cm]{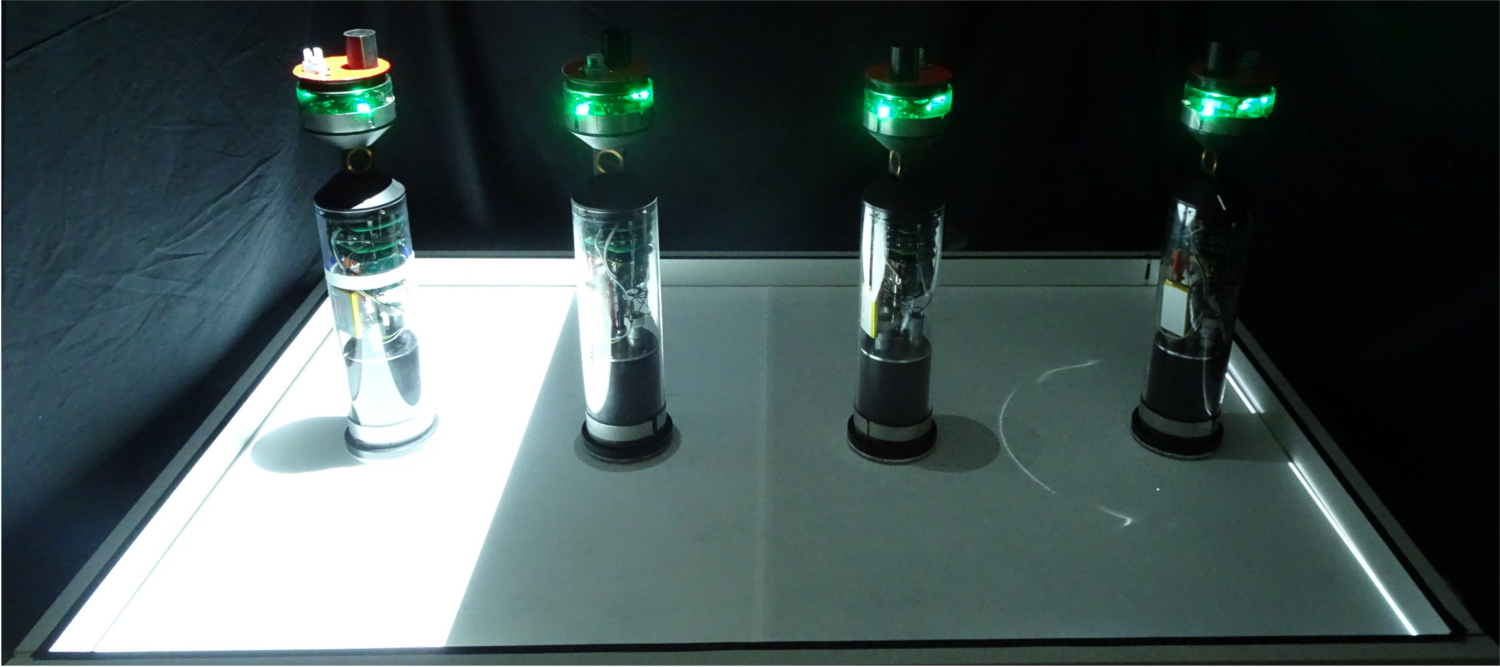}}
		\subfigure[]{\includegraphics[width=8cm]{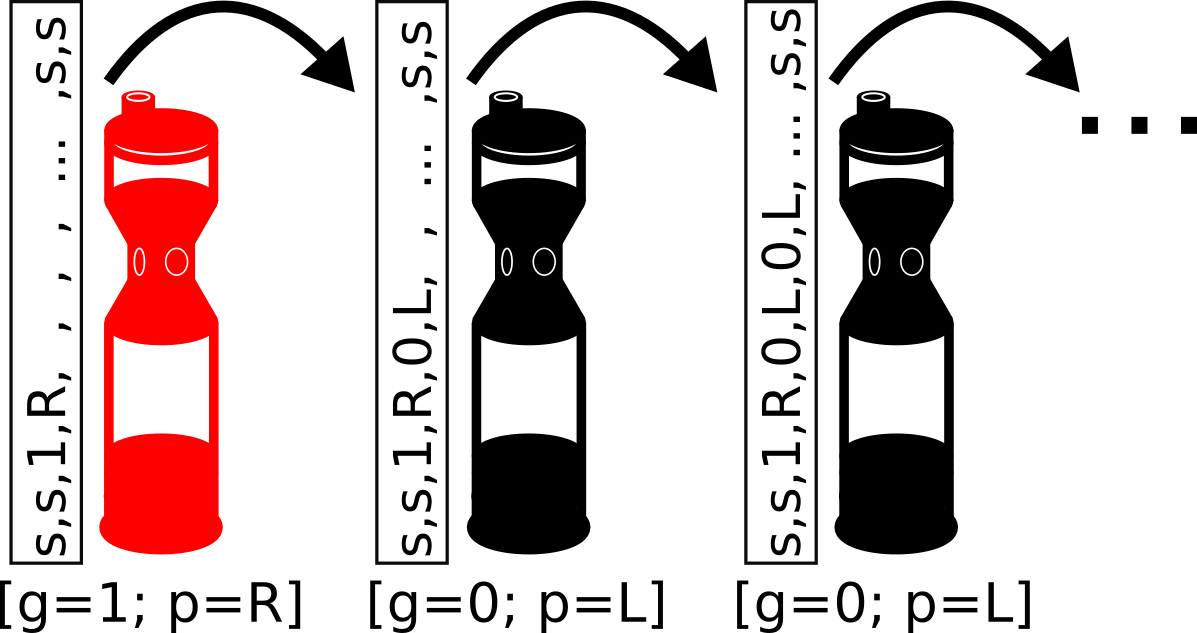}}
		\caption{(a) Four aMussels in an arena used for the experiments. (b) The experimental setup with one of the six considered light configurations. The aMussels did not decide on a preferred direction yet as their top caps are not illuminated.
		(c) An experiment counted as successful with all aMussels agreeing on moving left, towards the illuminated domain, indicated by the green LEDs in their top caps.
		(d) Schematic illustration of broadcasting and relaying messages. The red colored aMussel initiates a message containing its own local ambient light measurement value (here: $g\in\{0,1\}$) as well as its own preferred direction ($p\in\{R,L\}$). Other aMussels which successively receive the message (shown on the right side in black color) add their own sensor readings and preferred directions to the message before relaying it, respectively. 
		The messages broadcasted by the aMussels are shown to the left of each aMussel. Messages started and ended with the characters ``ss'' for easier parsing.}
      \label{fig:experiment}
\end{figure}

For testing the algorithm we used aMussels under lab conditions outside water in an one-dimensional setup. Four aMussels were arranged in a linear manner in an arena as shown in Figure~\ref{fig:experiment} (a).
As an emulation of oxygen gradients we used an ambient light gradient which allowed us to perform the experiments in the lab outside of a water environment. We hence were able to establish precisely controlled environmental situations and predictably changing environments. 
Two projectors were located above the arena and used for varying the light intensity on the arena floor as shown in Figure~\ref{fig:experiment}(b) and (c) where different parts of the floor are brightly illuminated and others dark. In this experiments we considered two states of illuminance: lights on or off. The setup of the system corresponds to the simulation of a swarm close to a sharp transition, presented in Section~\ref{sec:simulation_sharp}. The sensor for measuring ambient light values is located at the top cap of the aMussels. In this experiment they communicated via modulated green light. We counted an experiment as successful as soon as the robots agreed on the direction towards the border between the two different domains. While lights for communication are located in the center of their body, the LED's in their top caps (as visible in Figure~\ref{fig:experiment}(c) in green) indicated their preferred direction. From the perspective of the camera, green represents the preferred direction ``left'' and blue represents ``right''. 

The algorithm introduced in Section~\ref{sec:algorithm} was implemented on the aMussels, however the information gathering and negotiation phases were fused into a single phase. All messages sent by aMussels in this experiment contained both the sensor readings as well as their preferred direction. This is illustrated in Figure~\ref{fig:experiment}(d), where the red aMussel initiates the sending of a message. For this, it broadcasted its own sensor reading as well as its current preferred direction as message. When another aMussel received a message, it stored it and afterwards appended its own sensor reading and current preferred direction to this message before relaying it.\\
Based on the sensor readings in the stored messages, the aMussels continuously evaluated from which direction they received messages with largest variance in measurements, i.e. which direction they individually considered most preferable to move towards and likewise which direction they broadcasted as their own current preferred direction.\\
Based on the preferred directions in the stored messages, at the end of this phase (consisting of both the information gathering phase and the negotiation phase) the aMussels evaluated which direction was favored by the majority of the swarm and thus which direction they ultimately decided on moving towards.
This phase consisted of a time period of 10 cycles, meaning every aMussel initiated at least 10 messages. \\
The parameter values used in the experiment are given in Table~\ref{tab:table_params}. The robots randomized the time when they initiated a message during a cycle at the beginning of every cycle. As a result in this experiment the effective cycle length of individual robots varied between 40 and 70 seconds as indicated in Table~\ref{tab:table_params}. The reason for this randomization is that occasionally robots initiated message approximately at the same time. In this case the messages were not received by all other robots since right after broadcasting a message every robot stays insensitive to incoming messages for a brief amount of time. Due to randomizing the initiation of messages this event less likely occurred repeatedly.

\begin{table}[]
\centering
\begin{tabular}{l|l}
Parameter          & Value            \\ \hline \hline
Cycle length       & 55 +/- 15 seconds \\ \hline
Refractory time    & 4 seconds         \\ \hline
Message length     & 32 bytes          \\ \hline
Negotiation period & 10 cycle lengths \\ \hline
Length of messages & 32 byte
\end{tabular}
		\caption{The parameters values used in the experiments.}
      \label{tab:table_params}
\end{table}

\subsection{Experimental results}

\begin{figure}[thpb]
		\centering
		\subfigure[]{\includegraphics[width=2.6cm]{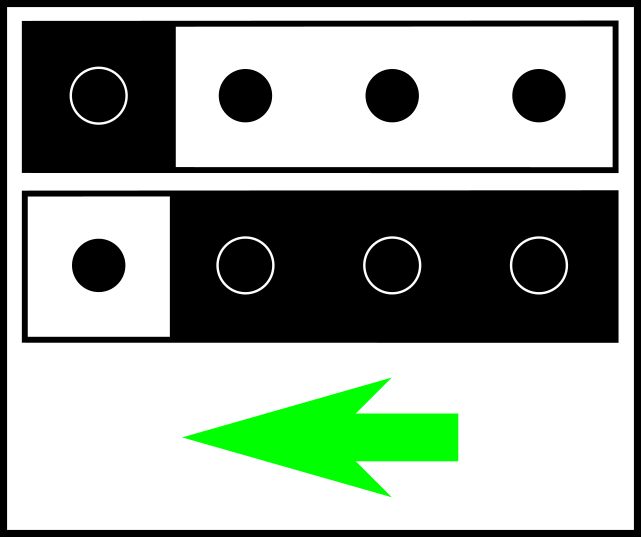}} \hfill
		\subfigure[]{\includegraphics[width=2.6cm]{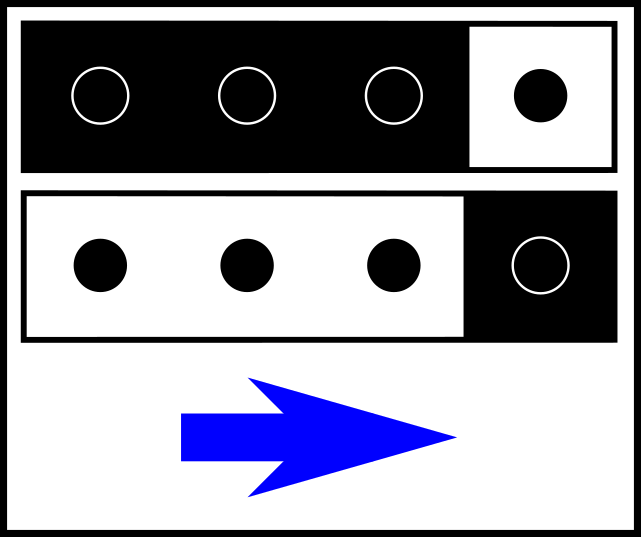}}\hfill	
		\subfigure[]{\includegraphics[width=2.6cm]{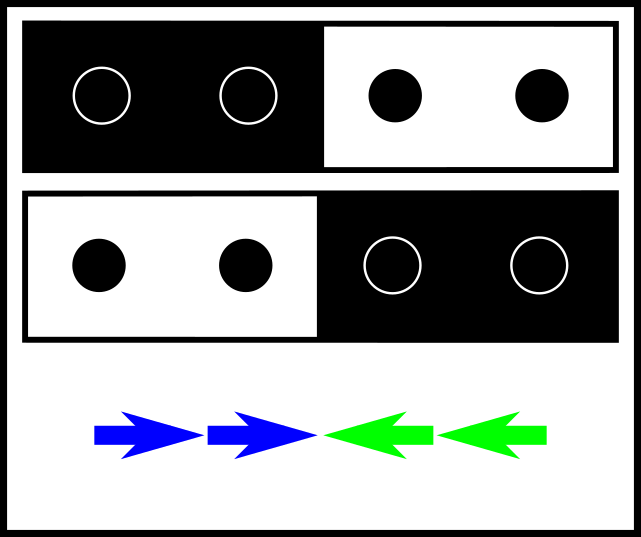}}
		\caption{Six different light configurations which were tested in the experiments. 
		(a) and (b) show the configurations for which we expect aMussels to decide to move to the left and to the right, respectively. The color of the arrow denotes the corresponding color the aMussels used to indicate their preferred direction via the LEDs in their top caps.
		(c) The configurations for which we expect the aMussels to not agree on a common direction but to choose directions towards the border of the domains of different luminosity. The two aMussels left from the border decide to move to the right and vice versa.}
      \label{fig:experiment_inkscape}
\end{figure}

We conducted experiments for six different light configurations as schematically shown in Figure~\ref{fig:experiment_inkscape}. The arrows to the right of each configuration indicate the results of each set of experiments. The direction of the arrow indicates the collective decision of the swarm in which direction to move and its color denotes the color used by the aMussels to indicate the respective direction they decided on (e.g. see Figure~\ref{fig:experiment}(c)). For Figure~\ref{fig:experiment_inkscape}(c) we counted an experiment as successful if the aMussels ultimately decided to move towards the border between the two domains of luminosity, i.e. the two aMussels on the left choose to move towards the right and vice versa.

For each light configuration the experiment was independently repeated five times with a success rate of 100 \%.
In order to test how well the aMussels adapt to changing light configurations we conducted another set of experiments with alternating light configuration in which the robots need to change their previously reached consensus. After reaching consensus, the light configurations were changed such that the expected direction for the robots to decide on was inverted. The experiments were considered as successful if the robots correctly found consensus for the initial light configuration and then switched their opinion accordingly. We conducted this experiment five independent times with all experiments successful.

\section{Discussion}
\label{sec:discussion}
In this paper, we demonstrated how a simple bio-inspired communication behavior can be used to reach a swarm level decision of which direction to move in order to maximize swarm level information access. We also demonstrated how this algorithm works in robots of an underwater swarm with limited communication range and local information. 
We presented simulation results in Section~\ref{sec:simulation} to give an intuitive understanding of the algorithm's functionality. 
For both a spatially discrete as well as a gradual change in measured quantity $X$ in the system the swarm could successfully maximize its diversity in the measurement.
For system with a discrete change in $X$ (Section~\ref{sec:simulation_sharp}) the swarm, within proximity of nearby variations, succeeded in $100\%$ of all simulations whereas for systems with a gradual change in $X$ (Section~\ref{sec:simulation_gradient}) the success rates vary between $85\%$ to $100\%$. Also the mean success times in the latter system (Figure~\ref{fig:smooth_transition_analysis}) are significantly larger compared to the prior (Figure~\ref{fig:discrete_transition_analysis}). It shows that the artificial swarms using this algorithm perform better the steeper a gradient in measured quantity is in the system.\\
It is also worth pointing out that the configuration of a swarm, i.e. the spatial distribution of agents, has a significant influence on the preferred direction. Considering a system of entirely random values in $X$. If agents are distributed e.g. in a line, the preferred direction can only be along the linear distribution of the swarm as messages are only shared along this line. A swarm shaped as a perfect cross will theoretically move like the king on a chess board, solely up-down-left-right. This needs to be taken into account in case a swarm tends to group or shape up in symmetric ways in order to avoid systematic errors. 

In Section~\ref{sec:experiments} we presented experimental results of a simplified laboratory demonstration of the algorithm implemented on robots using $N=4$ aMussels in a one-dimensional setup. 
The resulting behavior is in full qualitative agreement with the results of the corresponding numerical simulations (Figure~\ref{fig:discrete_transition_vectorfield}) in $100\%$ of the experiments. The chances of reaching an indecision point, when two aMussels decide to move left and two aMussels decide to move right (Figure \ref{fig:experiment_inkscape}(c)), decrease with increasing number of members of a swarm --- for a sufficiently large swarm of robots distributed in two spatial dimensions those chances of reaching an indecision point would be negligibly small. Despite the simplification of restricting the experiments to one dimension, they serve as proof of concept and general functionality of the implementation of the algorithm in robots.

For both simulations and experiments shown in this work we assumed/ensured an interconnected swarm with every agent being connected to at least one neighbor at all times. This assumption allowed the demonstration of the collective decision aspect of the algorithm, however it is not a feasible assumption in a real world scenario. Although it could be shown in previous work how the underlying communication mechanism exhibits significant resilience to signal loss~\citep{varughese2017b}, in practice a number of steps need to be taken in order to ensure connectivity of robots. In a real world scenario, one has to account for possible occlusions, alignment problems and other prospective challenges while using modulated light communication. 

For evaluating incoming messages between robots we used the variance of the received measurements. Although variance is a simple measure, it is an effective measure of information entropy for a swarm measuring a single parameter. In contrast to our approach,~\cite{cui2004swarm} use a fuzzy logic based evaluation. Such complex measures could be used in place of variance in the CIMAX algorithm when dealing with complex parameter spaces while following the same information gathering, evaluation and collective decision phases.

Lastly, in this work we only considered a single quantity being measured, however this algorithm constitutes a general approach for collective decision making in this particular class of swarms or networks. Therefore several quantities can be considered, resulting in a swarm maximizing the data points within a phase space spanned by the number of considered quantities. This hence lets such a swarm autonomously explore an environment of high complexity, taking into account previously collected data and adjusting to environmental changes and variations.

As the algorithm could be proven conceptually functional with respect to collective decision making it will be implemented and tested in the future on larger swarms and in-field within the framework of subCULTron.

\section*{ACKNOWLEDGMENT}

This work was supported by EU-H2020 Project subCULTron, funded by the European Unions Horizon 2020 research and innovation programmer under grant agreement No 640967. Furthermore this work was supported by the COLIBRI initiative at the University of Graz.


\bibliographystyle{apalike}
\bibliography{newBib}
\end{document}